\documentclass[10pt]{article}
\usepackage{times}
\usepackage[pdftex]{graphicx}
\usepackage{algorithm}
\usepackage{algorithmic}
\usepackage{amsthm}
\usepackage{tabularx}
\usepackage{multicol}
\usepackage{multirow}

\newtheorem{lemma}{Lemma}
\title{Finding All Bounded-Length Simple Cycles in a Directed Graph}
\author{Anshul Gupta and Toyotaro Suzumura \\
{\em anshul@us.ibm.com, suzumura@acm.org} \\
IBM T.J. Watson Research Center\\
Yorktown Heights, NY 10598, USA}
\date{\today}

\begin{document}

\maketitle

\begin{abstract}
A new efficient algorithm is presented for finding all simple cycles that satisfy a length constraint in a directed graph. When the number of vertices is non-trivial, most cycle-finding problems are of practical interest for sparse graphs only. We show that for a class of sparse graphs in which the vertex degrees are almost uniform, our algorithm can find all cycles of length less than or equal to $k$ in $O((c+n)(k-1)d^k)$ steps, where $n$ is the number of vertices, $c$ is the total number of cycles discovered, $d$ is the average degree of the graph's vertices, and $k > 1$. While our analysis for the running time addresses only a class of sparse graphs, we provide empirical and experimental evidence of the efficiency of the algorithm for general sparse graphs. This algorithm is a significant improvement over the only other deterministic algorithm for this problem known to us; it also lends itself to massive parallelism. Experimental results of a serial implementation on some large real-world graphs are presented.   
\end{abstract}


\section{Introduction}
\label{sec-intro}

Finding all simple cycles in a directed graph is a classical computer science 
problem, for which an elegant algorithm was published by 
Johnson~\cite{JOHNSON75sicomp}. An optimal and asymptotically faster algorithm 
for undirected graphs was proposed by 
Birmel\'{e}~{\em et al.}~\cite{BIRMELE2013soda}.

The number of cycles in a complete graph grows 
exponentially with the number of vertices, $n$. Even sparse graphs, 
such as planar graphs, can have an exponential number of 
cycles~\cite{BUCHIN07}. A more tractable problem that arises in real 
applications is that of finding all simple cycles of a given length $k$ 
or of length bounded by $k$, where $k$ is typically a small integer. 
This has applications in the analysis of social networks, communications 
graphs, and financial transaction graphs, etc. Despite its several 
applications, to the best of our knowledge, no deterministic algorithm for 
specifically this problem has been proposed to date. 
The closest problem that has been studied is that of finding $c$ shortest
simple cycles by Agarwal and 
Ramachandran~\cite{Agarwal2016isaac,Agarwal2018stoc}. Their $O(c n e)$ 
algorithm for enumerating the first $c$ cycles in nondecreasing order in a 
graph with $n$ vertrices and $e$ edges can be easily adapted to 
find all cycle with at most $k$ edges by stopping as soon as the
first cycle of length greater than $k$ is found.
Monien~\cite{MONIEN85} gave an $O(n e)$ algorithm for 
finding and reporting a cycle of length $k$ for any $k \geq 3$, 
where $n$ is the number of vertices and $e$ is the number of edges in 
the graph. Alon {\em et~al.}~\cite{ALON97algorithmica} presented an algorithm 
that improved this bound to $O(e^{2-2/k})$, reducing the running time 
for small values of $k$. 

In this paper, we present a relatively simple, but fast and powerful new algorithm for finding all simple cycles of length less than or equal to $k$ in sparse directed graphs. We show that for graphs with a nearly uniform vertex degree $d$, our algorithm takes $O((c+n)(k-1)d^k)$ time, where $c$ is the total number of cycles discovered, $d = e/n$, and $k > 1$. While our analysis for the running time addresses only a class of sparse graphs, we provide empirical and experimental evidence of the efficiency of the algorithm for general sparse graphs as well. For the class of graphs analyzed, the running time increases linearly with the size of the graph or the number of cycles discovered because for a given $k$ and a given graph, $kd^k$ is a constant, albeit relatively large. The algorithm is not only practical to implement, but also lends itself to easy and scalable parallelization, thus extending its suitability for real applications. 
%One such application considered in our experiments is the problem of detecting money-laundering patterns in financial transaction graphs.

The remainder of the paper is organized as follows. We start with an overview of Johnson's algorithm~\cite{JOHNSON75sicomp} for finding all simple cycles in Section~\ref{sec-johnson}, since our algorithm uses a similar overall strategy to avoid futile searches. We describe our algorithm in detail in Section~\ref{sec-new-algo}. Sections~\ref{sec-correct} and~\ref{sec-complexity} contain correctness proofs and complexity analysis, respectively. Experimental results on large graphs related to real applications are presented in Section~\ref{sec-experiment}. Finally, Section~\ref{sec-conc} contains concluding remarks and directions for future work.

\section{Johnson's algorithm for finding all simple cycles}
\label{sec-johnson}

\sloppy
Johnson's algorithm~\cite{JOHNSON75sicomp} combines depth-first search (DFS) and backtracking with an astute strategy of blocking and unblocking vertices to prevent fruitless searches and can find all simple cycles in a directed graph in $O((c+1)(n+e))$ time. It remains the best-known algorithm for this problem to date. At the time of its publication, it was a significant improvement over the previous fastest algorithm~\cite{TARJAN73siam}, whose worst-case complexity is $O(e.n(c+1))$. 

In each of its steps, Johnson's algorithm searches for cycles that start and end at a vertex $s$, in the strongly connected component~\cite{TARJAN72siam} of the subgraph induced by vertices in $\{s, s+1, \ldots, n\}$. Thus, step $s$ outputs all cycles in which $s$ is the vertex with the smallest index (henceforth, referred to as the ''smallest vertex"). Such cycle searches are performed for all values of $s$ from $1$ to $n-1$ for which the strongly connected component has more than $1$ vertex.

As the algorithm proceeds with DFS starting at vertex $s$, it stacks and blocks the vertices encountered along the way. The search would either lead to $s$, in which case a cycle has been discovered, or lead to a blocked vertex. Backtracking from an unsuccessful search unstacks the vertices, but leaves them blocked. It also adds the current vertex to the Blist of the most recently unstacked vertex, thus leaving a backward trail along an unsuccessful path. Blists are data structures used to remember unsuccessful portions of a path that has been explored. The blocked vertices prevent the search from following the same unsuccessful path, or a portion of it, more than once while the vertices responsible for the failure of the path are still on the stack. When backtracking from a successful path, the algorithm unblocks each unstacked vertex, and recursively unblocks the vertices in the Blist of each vertex being unblocked. The rationale is that a vertex $v$ being unstacked was the reason for blocking the vertices in the backward trails emanating from this vertex. As soon as $v$ leaves the stack after a successful search, all the vertices that it was responsible for blocking can be considered for inclusion in a cycle in a subsequent visit via a DFS path that reaches $v$ with a different stack. For a more detailed description and analysis of the algorithm, the reader is referred to the original paper~\cite{JOHNSON75sicomp}.

\section{Algorithm for length-constrained simple cycles}
\label{sec-new-algo}

We now present an efficient algorithm for finding all elementary cycles of length less than or equal to $k$ in a directed graph $G = (V,E)$ where $V \in \{1, 2, \ldots, n\}$ and $(u,v) \in E$ iff there is an edge with source $u \in V$ and destination $v \in V$.

\begin{figure}
\begin{tabbing}
xxx \= xx \= xx \= xx \= xx \= xx \= \kill
1. \> function LC\_CYCLES ($G$, $k$) \\
   \> // $G = (V, E)$, $V = \{1, 2, \ldots, n\}$ \\
2. \>\> for $s$ = 1, $n$ \\
3. \>\>\> $H^s = (W^s, F^s)$, where $W^s = \{s, s+1, \ldots, n\}$ and \\
   \>\>\>\>\> $(u,v) \in F^s$ if $u,v \geq s$ and $(u,v) \in E$; \\
4. \>\>\> $G^s = (V^s, E^s)$, where $V^s \in W^s$ is the set of all vertices reachable from $s$ \\
   \>\>\>\>\> via BFS with $k-1$ levels and $(u,v) \in E^s$ if $u$, $v \in V^s$, $(u,v) \in F^s$; \\ 
5. \>\>\> for $v \in V^s$ \\
6. \>\>\>\> Lock($v$) = $\infty$; \\
7. \>\>\>\> Blist($v$) = $\emptyset$; \\
8. \>\>\> end for \\
9. \>\>\> {\em blen} = CYCLE\_SEARCH ($G^s$, $s$, $k$, 0); \\
10.\>\> end for \\
11.\> end LC\_CYCLES
\end{tabbing}
\caption{\noindent Outline of the LC\_CYCLES procedure for finding all simple cycles of length less than or equal to $k$ in a directed graph $G = (V,E)$.}
\label{fig-algo-outer}
\end{figure}

\subsection{The outer loop}

Figure~\ref{fig-algo-outer} shows the outer loop of our algorithm. Similar to Johnson's algorithm, our algorithm also searches for cycles whose smallest vertex is $s$ in successive subgraphs $G^s$ for $s \in \{1, 2, \ldots, n\}$. However, the subgraph $G^s$ is constructed quite differently. Let $H^s$ be a subgraph of $G$ induced by $\{s, s+1, \ldots, n\}$. Then $G^s$ is the subgraph of $H^s$ induced by all the vertices in $H^s$ that are reachable by breath-first search (BFS) on $H^s$ up to $k-1$ levels starting with vertex $s$. Clearly, any vertex in $H^s$ that is not in $G^s$ cannot belong to a cycle that includes $s$ and has at most $k$ edges.  

\begin{figure}
\begin{tabbing}
xxx \= xx \= xx \= xx \= xx \= xx \= \kill
1. \> integer function CYCLE\_SEARCH ($G$, $v$, $k$, {\em flen}) \\
   \> // $G = (V, E)$, $V \in \{s, s+1, \ldots, n\}$ \\
2. \>\> {\em blen} = $\infty$; \\
3. \>\> Lock($v$) = {\em flen}; \\
4. \>\> PUSH\_STACK ($v$); \\
5. \>\> for $w \in$ Adjacency($v$) \\
6. \>\>\> if ($w == s)$ then \\
7. \>\>\>\> Output (Stack $\cup \{s\}$) as a new cycle of length {\em flen}$+1$; \\
8. \>\>\>\> {\em blen} = 1; \\
9. \>\>\> else if (({\em flen}$+1) <$ Lock($w$) and ({\em flen}$+1) < k$) then \\
10.\>\>\>\> {\em blen} = MIN ({\em blen}, 1 + CYCLE\_SEARCH ($G$, $w$, $k$, {\em flen}$+1$)); \\
11.\>\>\> end if \\
12.\>\> end for \\
13.\>\> if ({\em blen} $< \infty$) then \ \ \ \ \ \ \ \ \ \ \ \ \ \ \ \ \ \ \ // True only if at least one cycle found\\
14.\>\>\> RELAX\_LOCKS ($v$, $k$, {\em blen}); \\
15.\>\> else \\
16.\>\>\> for $w \in$ Adjacency($v$) if ($v \notin$ Blist($w$)) then Blist($w$) = Blist($w$) $\cup\ \{v\}$; \\ 
17.\>\> end if \\
18.\>\> POP\_STACK ($v$); \\
19.\>\> return ({\em blen}); \\
20.\> end CYCLE\_SEARCH
\end{tabbing}
\caption{\noindent The CYCLE\_SEARCH algorithm for finding all simple cycles of length at most $k$ that include vertex $s$ in a directed graph $G = (V,E)$ whose smallest vertex is $s$. Note that the forward distance from $s$, {\em flen}, is incremented at each recursive call.}
\label{fig-algo-search}
\end{figure}

\subsection{Cycle search with a given origin}

At the heart of our algorithm is the function CYCLE\_SEARCH shown in Figure~\ref{fig-algo-search} for finding all simple cycles of length at most $k$ that originate and terminate at vertex $s$ in subgraph $G^s$. When called to explore vertex $v$, the input parameter {\em flen} is the length of the current forward path from $s$ to $v$. The function returns the length of the shortest path from $v$ to $s$ when the current visit to $v$ results in the detection of a cycle; otherwise, it returns $\infty$.

In addition to a stack, this function uses two other key data structures. 

The first one is the array Lock, which stores an integer value for each vertex and helps avoid futile searches. For any vertex $v$ that has been reached at least once and is not currently on the stack, Lock($v$) contains one plus the maximum possible length of any path from $s$ to $v$ that has the possibility of reaching back to $s$ with at most $k$ total edges in it. This value is used to permit only those paths that reach $v$ with {\em flen} $<$ Lock($v$) to visit it and to block all other paths (Line~9). Note that Line~9 also blocks any path that already has $k-1$ edges and the next edge does not lead to $s$. As a path is being explored, Lock($v$) is set to {\em flen} when the path reaches $v$. If this path is eventually unsuccessful, then this value of Lock indicates that a future path reaching $v$ has the potential to succeed with its current stack only if it reaches $v$ with a smaller {\em flen} value. A cycles is detected when an edge $(v,s)$ is encountered (Line~6). At this point, {\em blen} or the backward distance from $s$ is set to 1 (Line~8). Upon return from every successful call to CYCLE\_SEARCH, {\em blen} at the calling vertex is updated (Line~10) so that the shortest backward distance from $s$ for the current vertex $v$ can be returned.

The second key data structure of function CYCLE\_SEARCH is the Blist associated with each vertex $w$, which stores the indices of the source vertices $v$ of all incoming edges $(v,w)$ for which no cycle containing $(v,w)$ is found with the current state of the search stack (Line~16). It can be implemented using a data structure such as B-trees~\cite{CORMENbook}, which permit fast insertion and look-up. Recall that the role of Lock($v$) is to allow vertex $v$ to be included only in those paths from $s$ that have the potential to terminate at $s$ within $k$ or fewer total hops (edges). When the most recent visit to $v$ results in an unsuccessful search, this is achieved by setting and leaving the value of Lock($v$) equal to {\em flen}. When $v$ is part of a successful search, then, with the help of the Blist data structures, the function RELAX\_LOCKS, called at Line~14 in CYCLE\_SEARCH, appropriately sets the Lock values of $v$ and the vertices that can reach it based on the length of the current shortest path from $v$ to $s$.

\begin{figure}
\begin{tabbing}
xxx \= xx \= xx \= xx \= xx \= xx \= \kill
1. \> function RELAX\_LOCKS ($u$, $k$, {\em blen}) \\
2. \>\> if (Lock($u$) $< k-${\em blen}+1) then \\
3. \>\>\> Lock($u$) = $k-${\em blen}+1; \\
4. \>\>\> for ($w \in$ Blist($u$)) \\
5. \>\>\>\> if ($w \notin$ Stack) RELAX\_LOCKS ($w$, $k$, {\em blen}+1); \\
6. \>\>\> end for \\
7. \>\> end if \\
8. \> end RELAX\_LOCKS
\end{tabbing}
\caption{\noindent The recursive procedure for relaxing the Lock values when backtracking along a path that is part of at least one cycle. Note that the backward distance from $s$, {\em blen}, is incremented at each recursive call.}
\label{fig-algo-relax}
\end{figure}

\subsection{Backtracking and relaxing Lock}

Figure~\ref{fig-algo-relax} shows the recursive procedure for reassigning the Lock values when a cycle is detected. This function is called for a vertex $u$ with a corresponding integer {\em blen}, which is the length of the currently shortest known path from $u$ to $s$. The value of Lock($u$) is increased to $k-${\em blen}$+1$, if possible (Line~3). An increase in Lock($u$) is equivalent to relaxing the conditions for $u$ to be included in a future search because this is permitted for only those paths from $s$ that reach $u$ with a length strictly less than Lock($u$). If {\em blen} is the length of the shortest path from $u$ to $s$, then a new path from $s$ to $u$ with length $k-${\em blen} or less can succeed. Relaxing Lock($u$) can also potentially relax Lock($w$) if edge $(w,u)$ was part of a previous unsuccessful search that led to $w$ being added to the Blist of $u$. This process of going back to relax Lock values proceeds recursively, adding 1 to {\em blen} for each backward step (Line~5). This backward traversal is gated by the condition on Line~2 and follows only those backward tracks along which the Lock values can be increased. 

\subsection{Performance gains through sorting and parallelism}
\label{subsec-parallel}

Note that the number of vertices and edges in subgraph $H^s$ in Line~3 of Function LC\_CYLCES in Figure~\ref{fig-algo-outer} monotonically decreases with $s$. Preprocessing graph $G$ to sort and renumber its vertices in the decreasing order of the sum of their incoming and outgoing edges would ensure that the density or the average degree (ratio of number of edges to the number of vertices) of $H^s$ also decreases monotonically with $s$. This can potentially improve performance in two ways. First, on an average, this is likely to reduce the size of the search space for constructing $G^s$, as well as the size of $G^s$. Second, this has the potential to reduce the overall amount of search in CYCLE\_SEARCH. Recall that for each vertex $s$, CYCLE\_SEARCH looks for cycles starting and ending in $s$ only. All other cycles that are traversed in the process are ignored. Processing denser subgraphs (presumably with more cycles) with a root vertex $s$ of greater connectivity earlier is likely to reduce the number of valid cycles that are traversed but whose reporting is delayed until these are traversed again in the right $G^s$. In Section~\ref{sec-experiment}, we show the improvement in running time achieved by this preprocessing step.

Function LC\_CYLCES can be easily and scalably parallelized, which significantly increases the algorithm's application potential for solving real-world problem involving very large graphs. All calls to CYCLE\_SEARCH on Line~9 are independent and can be performed in parallel. The amount of work in CYCLE\_SEARCH generally decreases as $s$ increases, especially if the vertices of $G$ are numbered in decreasing order of their degrees. In addition to the size of $G^s$, the amount of computation in CYCLE\_SEARCH also depends on the number of cycles that it detects; i.e., the number of cycles in which $s$ is the smallest vertex. Therefore, an efficient parallel implementation would need to carefully batch groups of starting vertices $s$ in order to avoid load imbalance. However, other that this relatively minor consideration, parallelization of LC\_CYLCES is quite straightforward.

\section{Correctness proof} 
\label{sec-correct}

The following three lemmas show that function CYCLE\_SEARCH in Figure~\ref{fig-algo-search} finds every cycle of length less than or equal to $k$ that includes vertex $s$ in $G^s$ exactly once.

\begin{lemma}
\label{lemma-a}
Function CYCLE\_SEARCH does not output any cycle of length greater than $k$.
\end{lemma}

\begin{proof}
By construction, in CYCLE\_SEARCH($G$,$v$,$k$,{\em flen}), {\em flen} is the length of the path from $s$ to $v$. Line~9 ensures that {\em flen} $< k$. Therefore, if an edge $(v,s)$ is encountered on Line~6, the length of the cycle is {\em flen}$+1$, which is $\leq k$.
\end{proof}

\begin{lemma}
\label{lemma-b}
Function CYCLE\_SEARCH outputs every cycle of length $k$ or less in subgraph $G^s$.
\end{lemma}

\begin{proof}
Assume that Lemma~\ref{lemma-b} is false, so CYCLE\_SEARCH never reaches Line~7 with $v = v_l$ and stack $(s,v_1,v_2,\ldots,v_l)$ for some valid cycle $(s,v_1,v_2,\ldots,v_l,s)$ in $G^s$. Without loss of generality, also assume that among the valid cycles for which this lemma fails, $(s,v_1,v_2,\ldots,v_l,s)$ is the first one. 

Since $(s,v_1,v_2,\ldots,v_l,s)$ is a valid cycle of length $l+1$, 
\begin{equation}
\label{eq-1}
l+1 \leq k.
\end{equation}
Since $(s,v_1,v_2,\ldots,v_l,s)$ is a cycle in $G^s$, $s$ must be in the adjacency list of $v_l$ and if CYCLE\_SEARCH($G^s$,$v_l$,$k$,$l$) reaches Line~6 with stack $(s,v_1,v_2,\ldots,v_l)$, the cycle is guaranteed to be detected. Therefore, for Lemma~\ref{lemma-b} to fail, the algorithm must be blocked at some edge other than $(v_l,s)$ in this cycle. Without loss of generality, assume that $(v_i,v_{i+1})$ is the first blocked edge of this cycle, where $v_0 = s$ and $0 \leq i < l$. 

Since $i < l < k$, edge $(v_i,v_{i+1})$ can be blocked only if

\begin{equation}
\label{eq-2}
i+1 \geq \mbox{Lock}(v_{i+1}) 
\end{equation}
on Line~9 because {\em flen} for $v_i$ is $i$. This cannot happen if this is the first visit to $v_{i+1}$ because all Lock values were initialized to $\infty$ in LC\_CYCLES before the call to CYCLE\_SEARCH. There are only two other possibilities for Inequality~\ref{eq-2} to be true.

The first possibility is that the last visit to $v_{i+1}$, which would have happened with a stack different from the current one, did not yield a valid cycle. In this case, Lock($v_{i+1}$) would have been set equal to {\em flen} for that visit. Let's denote that {\em flen} with $f$. By Inequality~\ref{eq-2}, it follows that 
\begin{equation}
f \leq i+1. 
\label{eq-3}
\end{equation}
Since $(s,v_1,v_2,\ldots,v_l,s)$ is a cycle, we know that a path from $v_{i+1}$ to $s$ of length $l-i$ exists. Aslo, since $v_{i+1}$ was visited earlier with a different stack, a cycle other than $(s,v_1,v_2,\ldots,v_l,s)$ containing $v_{i+1}$ exists. The length $q$ of this cycle is $f+l-i$, or $f = q-l+i$. By Inequality~\ref{eq-3}, $q-l+i \leq i+1$ or $q \leq l+1$. Therefore, by Inequality~\ref{eq-1}, $q \leq k$. This means that this other cycle containing $v_{i+1}$ meets the length constraint and is a valid cycle. Since $(s,v_1,v_2,\ldots,v_l,s)$ is the first cycle to be missed by the algorithm, the previous visit to $v_{i+1}$, which set Lock($v_{i+1}$) to a value less than or equal to $i+1$, must have produced a cycle of length $q$. This leads to a contradiction because we are considering the scenario in which the last visit to $v_{i+1}$ did not yield a valid cycle. This rules out the first possibility.

The second possibility is that the last visit to $v_{i+1}$ did yield a valid cycle, and therefore, function RELAX\_LOCKS (Figure~\ref{fig-algo-relax}) was called with $v_{i+1}$ as the first argument. Let $b$ be the value of {\em blen} in that call to RELAX\_LOCKS. By design of RELAX\_LOCKS, the following must hold upon return from that call:
\begin{equation}
\mbox{Lock}(v_{i+1}) \geq k-b+1.
\label{eq-4}
\end{equation}
By virtue of $(s,v_1,v_2,\ldots,v_l,s)$ being a cycle, there is path of length $l-i$ from $v_{i+1}$ to $s$. This path must have been successfully traversed during the last visit to $v_{i+1}$ because $(v_i,v_{i+1})$ is the first edge of any valid cycle to be blocked. Therefore, Line~10 in the last call to CYCLE\_SEARCH($G^s$,$v_{i+1}$,$k$,{\em flen}) will ensure that
\begin{equation}
b \leq l-i.
\label{eq-5}
\end{equation}
From Inequalities~\ref{eq-2} and~\ref{eq-4}, it follows that $i+1 \geq k-b+1$ or $b \geq k-i$. Combining this with Inequality~\ref{eq-5} yields $k \leq l$, which contradicts Inequality~\ref{eq-1}. Therefore, the second possibility is also ruled out.

In conclusion, no edge of the cycle $(s,v_1,v_2,\ldots,v_l,s)$ can be blocked in CYCLE\_SEARCH if $l < k$, and this cycle is guaranteed to be reported by the algorithm. This completes the proof. 

\end{proof}

\begin{lemma}
\label{lemma-c}
Function CYCLE\_SEARCH does not output any cycle $(s,v_1,v_2,\ldots,v_l,s)$, where $l < k$, more than once.
\end{lemma}

\begin{proof}
Line~5 of CYCLE\_SEARCH visits each element in the adjacency list of $v$ exactly once. After edge $(v_l,s)$ is detected and $v_l$ is unstacked at the end of CYCLE\_SEARCH($G^s$,$v_l$,$k$,$l$), it can never get back to $v_l$ with the same stack again and the path $(s,v_1,v_2,\ldots,v_l)$ is never explored further again. 
\end{proof}

Lemmas~\ref{lemma-a}--\ref{lemma-c} prove that CYCLE\_SEARCH yields all valid cycles starting and ending at $s$ in $G^s$ exactly once. Since CYCLE\_SEARCH is called once for all $s \in \{1,2,\ldots,n\}$ in LC\_CYCLES, it can be concluded that LC\_CYCLES yields exactly one instance of every valid cycle in the original graph $G$.

\section{Time complexity}
\label{sec-complexity}

We first analyze the running time of CYCLE\_SEARCH for finding all cycles starting and ending at vertex $s$ in graph $G^s$. In CYCLE\_SEARCH, the Lock values monotonically decrease until a valid cycle is found. Additionally, a vertex cannot be revisited without decreasing its Lock value. This ensures that no vertex is visited more than $k-1$ times before a valid cycle is detected. After a cycle is found, the Lock values of some vertices are relaxed or increased. During the process of relaxing the Lock values, function RELAX\_LOCKS does not traverse an edge unless a Lock value can be increased. In the worst case scenario, this cannot happen to any vertex at most $k-1$ times. Therefore, the time associated with finding a valid cycle cannot exceed $O((k-1)(|V^s|+|E^s|)$, and the time complexity of CYCLE\_SEARCH for finding all valid cycles in a given subgraph $G^s$ is $O((c_s+1)(k-1)(|V^s|+|E^s|)$, where $c_s$ is the number of valid cycles in which the smallest vertex is $s$.

Recall that $G^s$ consists of $s$ and all vertices greater than $s$ that are reachable from $s$ within $k-1$ steps of BFS. If $G$ is a sparse graph with a uniform degree $d$, then $|V^s|+|E^s|$ is $O(d^k)$ for all $s$. Therefore, the complexity of the overall algorithm for finding all cycles of length $k$ or less in a graph with a uniform degree $d$ is $O(\sum_{s=1}^{n}(c_s+1)(k-1)d^k)$ = $O((c+n)(k-1)d^k)$, where $k > 1$ and $c = \sum_{s=1}^{n}c_s$ is the total number of valid cycles in $G$.

In our analysis, we use $d^k$ as the upper bound on the size of any $G^s$ in the call to CYCLE\_SEARCH. For sufficiently large graphs, the size of $G^s$ would not increase with $n$ as long as $d = e/n$ stays constant. $G^s$ consists of all vertices with a maximum distance of $k-1$ from $s$, and the size of this feasible neighborhood of $s$ depends only the structure of $G$ (unless $G$ is a scale-free network, which is rare~\cite{Broido-2019}). The simplifying assumption of uniform vertex degrees was made to arrive at a closed-form expression for an upper bound on the size of $G^s$. For sufficiently large $n$, the size of $G^s$ is likely to be independent of $n$ even without this assumption. As a result, for a given $k$ and a given $d$, the overall time of our algorithm would increase only linearly with the number of valid cycles and the number of vertices in the graph. In Section~\ref{sec-experiment}, we experimentally examine the growth of the running time with respect to $k$, and therefore with $c$, which itself is likely to grow with $k$.

It should be noted that explicit construction of $G^s$ is neither necessary for the correctness of the algorithm, nor does it have a significant effect on its time complexity. By construction, function CYCLE\_SEARCH does not visit any vertex whose minimum distance from $s$ is $k$ or greater. By using $G^s$ instead of $H^s$, CYCLE\_SEARCH avoids testing unnecessary edges at the periphery of the feasible neighborhood of $s$ on Line~9. Additionally, explicitly constructing $G^s$ with its own local indices improves the memory access locality of a computer implementation of the algorithm and yields better performance.

\section{Experimental results}
\label{sec-experiment}

\begin{table}
\begin{center}
\begin{tabular}{|l|c|c|c|c|} \hline
Graphs & $n$ & $e$ & $d$ & $\sigma_d$ \\\hline 
amazon0505 & 4.10e5 & 3.36e6 & 8.18 & 3.07 \\
as-caida & 2.65e4 & 1.07e5 & 4.03 & 33.4 \\
email-EuAll & 2.65e5 & 4.20e5 & 1.58 & 9.97 \\
soc-pokec & 1.63e6 & 3.06e7 & 18.8 & 32.0 \\
web-BerkStan & 6.85e5 & 7.60e6 & 11.1 & 16.3 \\
web-Google & 8.76e5 & 5.11e6 & 5.83 & 6.59 \\
wiki-topcats & 1.79e6 & 2.85e7 & 15.9 & 30.4 \\\hline
\end{tabular}
\end{center}
\caption{\noindent Description of the directed graphs used in experiments in this paper. As elsewhere in the paper, $n$ refers to the number of vertices, $e$ refers to the number of directed edges, $d$ is the average out-degree or the ratio $\frac{e}{n}$ of the graph, and $\sigma_d$ is the standard deviation of the out-degrees of the vertices.}
\label{table-datasets}
\end{table}

We present experimental results on a set of large directed graphs obtained from the SNAP data set~\cite{snapnets}. Table~\ref{table-datasets} contains detailed information about each graph. These graphs include social and communication networks, web graphs, product relationships, and professional networks, etc. They range in average density from roughly 1.6 to 18.8. The standard deviation of the out-degrees shows that with the exception of amazon0505, the graphs are fairly nonuniform. This is not surprising, since it has been observed that many social networks are highly skewed graphs and some even exhibit power-law characteristics~\cite{Barab_si_1999}. For the results reported in this section, a serial implementation of the algorithm in Figures~\ref{fig-algo-outer}--\ref{fig-algo-relax} was run on a single 3.3 GHz Intel Xeon E5-2667 core. This section contains results from our implementation only because we were unable to find any other practical algorithms or software to solve this problem for comparison on large graphs.

\begin{table}
\begin{center}
\begin{tabular}{|l|c|c|c|c|c|c|c|} \hline
\multirow{2}{4em}{Graphs} & \multicolumn{2}{|c|}{$k = 4$} & \multicolumn{3}{|c|}{$k = 5$} & \multicolumn{2}{|c|}{$k = 6$} \\\cline{2-8}
 & $c$ & $T$ & $c$ & $T$ & $T$ (uns) & $c$ & $T$ \\\hline
amazon0505 & 1.35e7 & 1.50 & 6.45e07 & 8 & 11 & 3.04e08 & 41 \\
as-caida & 4.65e6 & 3.67 & 1.47e08 & 252 & 401 & 7.64e09 & 11168 \\
email-EuAll & 6.68e6 & 5.0 & 2.95e08 & 407 & 472 & 1.50e10 & 27130 \\
soc-pokec & 5.21e8 & 1820 & 1.34e10 & 45606 & 69187 & 6.11e11 & 1.24e6 \\
web-BerkStan & 8.56e8 & 329 & 7.55e10 & 58702 & 62043 & - & - \\
web-Google & 3.39e7 & 3.35 & 3.11e08 & 44 & 97 & 3.35e09 & 638 \\
wiki-topcats & 1.93e8 & 521 & 6.36e09 & 17165 & 1.03e6 & 3.28e11 & 1.96e6 \\\hline
\end{tabular}
\end{center}
\caption{\noindent Total number of all cycles of length between 3 and $k$, for $k = 4, 5, 6$  and the time to find them in some directed graphs obtained from the SNAP repository~\cite{snapnets}. For $k = 5$, running time without sorting the vertices by out-degree is also included to show the benefit of sorting. The number of cycles is denoted by $c$, and the time $T$ is in seconds.} 
\label{table-snap}
\end{table}

Table~\ref{table-snap} shows the number of cycles detected and the running time of our algorithm for $k$ = 4, 5, and 6. While our algorithm detects all simple cycles of length less than or equal to $k$, we do not count and report cycles of length one (self-edges) and two (bidirectional edges) in this table. The table shows that the algorithm successfully detected up to hundreds of billions of cycles on large graphs using a single CPU, with only one run (web-BerkStan graph for $k = 6$) that did not complete in several days and had to be aborted. By default, we sort and renumber the vertices of the graph in increasing order of total degree, as discussed in Section~\ref{subsec-parallel}. For $k$ = 5, we also show the running time with this sorting turned off to show the benefit of sorting. The time is worse without sorting for every graph in our test set. In most cases, sorting results in an improvement by a relatively small factor; however, for one graph (wiki-topcats), the difference is so significant that the algorithm may not be practical without sorting.

\begin{table}
\begin{center}
\begin{tabular}{|l|c|c|c|} \hline
Graphs & $k = 4$ & $k = 5$ & $k = 6$ \\\hline
amazon0505 & 8.02e-12 & 8.40e-13 & 8.98e-14 \\
as-caida & 9.90e-10 & 4.03e-10 & 6.80e-11  \\
email-EuAll & 3.71e-08 & 3.45e-08 & 2.29e-08 \\
soc-pokec & 8.32e-12 & 3.67e-13 & 1.33e-14 \\
web-BerkStan & 8.46e-12 & 1.16e-12 & - \\
web-Google & 2.78e-11 & 5.24e-12 & 9.70e-13  \\
wiki-topcats & 1.39e-11 & 6.60e-13 & 7.36e-14  \\\hline
\end{tabular}
\end{center}
\caption{\noindent The ratio $\frac{T}{(c+n)(k-1)\mbox{min}(e,d^k)}$ for $k = 4, 5, 6$ for the graphs from the SNAP data set.} 
\label{table-ratio}
\end{table}

In Section~\ref{sec-complexity}, we derived $O((c+n)(k-1)d^k)$ as the expression for the time complexity of our algorithm for graphs with a uniform out-degree $d$. We tested the graphs in our test suite from the SNAP data set against this expression for growth of running time with $k$ and $c$. Table~\ref{table-ratio} shows the ratio $\frac{T}{(c+n)(k-1)\mbox{min}(e,d^k)}$ for $k = 4, 5, 6$ for these graphs. We used min$(e,d^k)$ in the denominator because $d^k$ is meant to represent the maximum number of edges in each $G^s$ in the CYCLE\_SEARCH algorithm (Figure~\ref{fig-algo-search}) for a given $k$ and in practice, it cannot exceed the total number of edges $e$ in the original graph. Table~\ref{table-ratio} shows that this ratio decreases as $k$ (and therefore, $c$) increase for all the graphs in our test set. Even though our graphs are quite diverse in origin and density, and the distribution of edges over the vertices is highly nonuniform, they all share this trend. The analytically derived running time bound used a simplifying assumption that the vertices had a uniform degree $d$. Experimentally observing that despite non-uniform degrees, the growth in the running time with $k$ and $c$ for every test case is slower than that suggested by the analytical bound is therefore an important finding and reinforces the practicality of our algorithm. 

\section{Concluding remarks and future work}
\label{sec-conc}

While enumerating simple cycles in directed graphs is a classical computer 
science problem, in most real applications on very large graphs, only a 
small fraction of the potentially enormous number of cycles are of interest. 
There has been relatively little research on discovering interesting cycles 
and algorithms for these are typically expensive~\cite{Adriaens19CIKM},
unless the search space can be substantially narrowed based on the definition
of interesting cycles, for example, cycles with obligatory vertices or
edges~\cite{KLAMT2009}. 
One class of cycle enumeration and detection problems of practical interest 
involves finding cycles of a limited length. We devised an efficient 
and practical algorithm for finding all length-constrained simple cycles 
in a directed graph.

We analytically and experimentally demonstrated the efficiency of the 
algorithm and showed that a serial implementation could detect billions of 
cycles on large graphs with millions on edges in a reasonable amount of time. 
In the future, it would be interesting to study the results of a parallel 
implementation on even larger graphs. On the analysis side, it would be 
interesting to explore the derivation of tighter time bounds without 
simplifying assumptions about the structure of the graphs.

\bibliography{mybib}

\end{document}